\documentclass[nofootinbib,twocolumn,amssymb]{revtex4}
\usepackage{verbatim}
\begin{document}

\title{Near Scale Invariance with Modified Dispersion Relations}

\author{C. Armendariz-Picon}
\email{armen@phy.syr.edu}
\affiliation{Physics Department, Syracuse University, Syracuse, NY 13244, USA.}

\date{\today}

\begin{abstract}
We describe a novel mechanism to seed a nearly scale invariant spectrum of adiabatic perturbations during a non-inflationary stage. It relies on a modified dispersion relation that contains higher powers of the spatial momentum of matter perturbations. We implement this idea in the context of a massless scalar field in an otherwise perfectly homogeneous universe. The couplings of the field to background scalars and tensors  give rise to the required modification of its dispersion relation, and the couplings of the scalar to matter result in an adiabatic primordial spectrum. This work is  meant to explicitly illustrate that it is possible to seed nearly scale invariant primordial spectra without inflation, within a conventional expansion history.

\end{abstract}

\maketitle

\section{Introduction}

Inflation has established itself as  a key idea in our understanding of the universe \cite{inflation}. A single stage of accelerated expansion can explain in one shot why the universe is homogeneous and isotropic, spatially flat, what is the origin of the entropy in the universe and why primordial perturbations are adiabatic, Gaussian and nearly scale invariant. For this reason, due to its central role in modern cosmology, it is important to ascertain whether there are alternatives to explain some of the inflationary predictions \cite{alternatives}, or whether the universe inevitably had to undergo an inflationary stage \cite{non-alternatives}.

The generation of an adiabatic, Gaussian and scale invariant spectrum of super-horizon perturbations arguably is the most important success of inflation. Indeed, one can reason that it is natural for the universe to begin in a symmetric state, thus accounting for its homogeneity and isotropy.   But in an expanding universe with a monotonically decreasing Hubble parameter, there seems to be no way to causally explain the origin of super-horizon perturbations other than inflation, regardless of their spectral index.

In previous works, Eugene Lim and I  explored alternative ways of seeding a scale invariant spectrum of primordial perturbations by relaxing some of the assumptions made in our descriptions of the origin of structure. In particular, in \cite{Armendariz-Picon:2003ht} we studied whether a contracting sound horizon can lead to the seeding of perturbations,  and in \cite{Armendariz-Picon:2003ku} I considered the impact of a time-varying  scalar field mass on the primordial spectrum. It turns out that in both cases, the generation of  scale invariant  super-horizon sized primordial perturbations still requires the existence of an inflationary stage, although some of the constraints on its properties can  be significantly relaxed. 

This work  is to some extent a natural continuation of these previous attempts. Our models on the origin of structure are based on extrapolations of low-energy theories to extremely high momenta. However, as we go back in time in the history of the universe, we expect physical laws to change as typical energies increase. In particular, the dispersion relations used to describe the evolution of structure should receive corrections suppressed by a cut-off scale $M$, which become important at high energies. A similar phenomenon occurs in crystal lattices; as the wave vector of an electron approaches a Bragg plane, its energy  deviates from the conventional quadratic behavior. 

In this work, we explore whether modified dispersion relations  containing higher powers of the spatial momentum $k$, can lead to exactly or nearly scale invariant primordial spectra  in a non-inflating universe.  Ironically, similar  dispersion relations  have been also  considered to study the robustness of inflationary predictions. Specifically, they have been used to analyze whether ``trans-Planckian" physics may significantly affect the spectrum of primordial perturbations seeded during inflation  \cite{Martin:2003kp}. Our goal here is quite different. We would like to determine whether modified dispersion relations may lead to nearly  scale invariant spectra in circumstances where we do not expect them to be so. 

\section{Modified Dispersion Relations}
Consider an expanding, homogeneous, isotropic and spatially flat universe, $ds^2=a^2(\eta)(-d\eta^2+d\vec{x}^2).$ Let us assume for simplicity that the universe power-law expands,
\begin{equation}\label{eq:a}
	a(\eta)=\left(\frac{\eta}{\eta_T}\right)^p,
\end{equation}
where we have arbitrarily set to one the value of the scale factor  at an arbitrary time $\eta_T$ we shall later define. The universe accelerates for $-\infty<p<-1$ and decelerates for $1/2<p<\infty$, where we have assumed that the dominant energy condition is satisfied. Concretely, a universe with effective equation of state parameter $w$ has 
\begin{equation}
	p=\frac{2}{1+3w}.
\end{equation}

In Fourier space, the equation of motion of a massless  scalar field  $\varphi$ in such a universe is 
\begin{equation}\label{eq:motion}
	v_k''+\left(\omega^2(k,\eta)-\frac{a''}{a}\right)v_k=0,
\end{equation}
where $v_k= a\cdot \varphi_k$, and a prime denotes a derivative with respect to conformal time.  In this work we assume that the dispersion relation of the scalar field is
\begin{equation}\label{eq:dispersion}
	\omega^2(k,\eta)=\frac{1 }{M^{2n-2}}\frac{k^{2n}}{a^{2m}},
\end{equation}
where $n$ and $m$ are constant parameters, and $M$ is a mass scale (recall that we have set $a_T=1$.)  Conventionally, the dispersion relation of a scalar field is $\omega^2=k^2$, that is, $n=1$ and $m=0$. Fluids with a non-constant speed of sound yield a dispersion relation with $n=1$ and $m\neq 0$. We discuss in Section \ref{sec:realizations} how dispersion relations with different values of $n$ and $m$ \cite{Brandenberger:2000wr} can arise from  generally covariant scalar field actions. In the meantime, the reader can think of equation (\ref{eq:dispersion}) as a phenomenological  description of new physics at momenta $k/a$ that at time $\eta_T$ are above the  cut-off $M$ \cite{Unruh:1994je,Corley:1996ar}. 

\subsection{Initial conditions}
The solution of equation (\ref{eq:motion}) with dispersion relation (\ref{eq:dispersion}) in a universe that expands according to equation (\ref{eq:a}) is
\begin{equation}\label{eq:solution}
	v_k(\eta)=\sqrt{\eta}\left[C_1\, H_\nu^{(1)}\left(\frac{\omega\cdot \eta}{mp-1}\right)+
	C_2 \,H_\nu ^{(2)}\left(\frac{\omega \cdot\eta}{mp-1}\right)\right],
\end{equation}
where the $H_\nu$ are the Hankel functions of the first and second kind,  and 
\begin{equation}
	\nu=\frac{1}{2}\left|\frac{2p-1}{mp-1}\right|.
\end{equation}
Note that the equation of motion  (\ref{eq:motion}) is invariant under $\eta\to -\eta$. Thus, in order to obtain a solution during inflation (where conformal time is negative), we can simply replace $\eta$ by $-\eta$. Because the differential  equation (\ref{eq:motion}) is second order, its general solution contains two arbitrary integration constants, $C_1$ and $C_2$. Their values are determined by requiring that  perturbations be in the adiabatic vacuum at early times. Essentially, this is the only way we know of to specify initial conditions.

The notion of an adiabatic vacuum exists only if the equation of motion admits the approximate solution
\begin{equation}\label{eq:vacuum}
	v_k\approx \frac{1}{\sqrt{2\,\omega(k,\eta)}} 
	\exp\left(-i\int^\eta \omega(k,\tilde{\eta}) \, d\tilde{\eta}\right),
\end{equation}
where $\omega$ is a possibly time-dependent frequency \cite{BirrellDavies}. This in fact resembles the vacuum mode functions of a field in Minkowski space. If such an approximate solution exists, perturbations are defined to be  in the adiabatic vacuum if they match the approximate solution (\ref{eq:vacuum}) at early times. 

It can be verified that in the limit $\omega \, \eta \gg 1$, the solution (\ref{eq:solution}) reproduces the adiabatic vacuum (\ref{eq:vacuum}) if
\begin{equation}\label{eq:initial}
	C_1=\frac{1}{2}\sqrt{\frac{\pi}{mp-1}}, \quad  C_2=0.
\end{equation}
Hence, what remains to be verified is whether $\omega\,\eta \to \infty$ at early times.  Substituting equation (\ref{eq:a}) into (\ref{eq:dispersion}) we find that $\omega\, \eta\gg 1$ amounts to
\begin{equation}
 	\left(\frac{H}{H_T}\right)^{\frac{mp-1}{p+1}} k^n \, H_T^{-1}\gg M^{n-1},
\end{equation}
where $H_T$ is the value of the Hubble parameter $H$ at the arbitrary time $\eta_T$. It is hence  clear that the adiabatic condition is satisfied at early times (when $H\to \infty$) if
\begin{equation}\label{eq:condition}
	(1+p) (mp-1)>0.
\end{equation}
Therefore, for a conventional dispersion relation, $n=1$ and $m=0$, we can specify vacuum initial conditions only when ${p<-1}$, that is, during inflation. However, if we drop this assumption, we find that an adiabatic solution at early times exists even if $p>1/2$, that is, even if the universe decelerates. Note that if condition (\ref{eq:condition}) is satisfied, $\omega^2\cdot \eta^2$ decreases as the universe expands. 

\subsection{Horizon Crossing}
It is generally argued that only during a stage of inflation it is  possible to seed primordial perturbations, regardless of their spectrum. The argument goes as follows. The physical length of a mode is $a/k$, and the Hubble radius is $H^{-1}$. Only during a stage of inflation does the length of a mode increase faster than the Hubble radius, that is, only during a stage of inflation can modes begin sub-horizon sized and later become super-horizon sized.   In other words, $k^2 \eta^2$ decreases only if the universe accelerates.

Our previous analysis shows the limits of the previous argument (see also \cite{Armendariz-Picon:2003ht}). Going back to the equation of motion (\ref{eq:motion}) one realizes that what really matters is whether $\omega^2$ grows faster or slower than than $a''/a$. Only if  $\omega^2\gg a''/a$ at early times there is a chance that we can find solutions that match the adiabatic vacuum (\ref{eq:vacuum}). We call modes for which $\omega^2>a''/a$ ``short-wavelength" modes, and modes for which $\omega^2<a''/a$ ``long-wavelength" modes. Typically, $a''/a \sim 1/\eta^2$, so a transition between the short and long-wavelength regimes occurs when $\omega^2 \eta^2$ decreases, irrespective of the universe expansion \cite{Armendariz-Picon:2003ht}. Moreover, in some (degenerate) cases even  this conclusion does not apply. In a radiation dominated universe, $a''/a$ equals zero, and hence all modes (both sub-horizon and super-horizon) are short-wavelength. 
 
Nevertheless, the existence of a short-wavelength regime is not sufficient for the existence of an adiabatic regime, which is the regime needed to specify meaningful initial conditions. Equation (\ref{eq:vacuum}) typically is an approximate solution of the equations of motion  only if the adiabatic condition
$\omega'/\omega^2\ll 1$ holds \cite{BirrellDavies}. This equation is formally different from the short-wavelength condition $\omega^2\gg a''/a$, although in practice both are equivalent. Because $\omega'/\omega \sim 1/\eta$ and $a''/a\sim 1/\eta^2$,   short-wavelength modes are generally adiabatic. Therefore, the only condition we have to impose on any  model for the origin of structure is  that $\omega^2 \, \eta^2$ be monotonically decreasing. Since the period of oscillation of the perturbations (in conformal time) is $1/\omega$, what this means is that the crucial requirement for the causal generation of perturbations is freeze-out ($\omega\, \eta > 1\to\omega\, \eta < 1$), rather than horizon-crossing ($k\, \eta>1 \to k \, \eta< 1$).

\subsection{The Power Spectrum}
The solution (\ref{eq:solution}) of the equation of motion (\ref{eq:motion}) with adiabatic vacuum initial conditions (\ref{eq:initial}) is
\begin{equation}\label{eq:exact}
	v_k=\frac{1}{2}\sqrt{\frac{\pi \eta}{mp-1}} H^{(1)}_\nu\left(\omega(k,\eta)\cdot \eta\right).
\end{equation} 
With this exact solution  in hand, we are ready to calculate the primordial spectrum of field perturbations. Because, by assumption, $\varphi$ is in a vacuum state, its expectation value vanishes, $\langle \varphi \rangle=0$. The two-point function of the field is characterized by the power spectrum
\begin{equation}
	\mathcal{P}_{\varphi}=\frac{k^3}{2\pi^2} |\varphi_k|^2,
\end{equation}
which is also a measure of the mean squared inhomogeneities of the field on comoving length-scales $1/k.$ Using the relation $v_k= a\, \varphi_k$ and equation (\ref{eq:exact}) 
we find that at late times ($\omega \, \eta\ll1$)  the Fourier mode $\varphi_k$ remains constant for ${\text{sgn}[(2p-1)/(mp-1)]=1}$.  In this case, the spectrum is given by 
\begin{equation}\label{eq:power}
	\mathcal{P}_\varphi= A^2 \cdot
	\left(\frac{k}{k_T}\right)^{n_s-1},
\end{equation}
where the spectral index is 
\begin{equation}\label{eq:index}
	n_s-1= 3-n \left|\frac{2p-1}{mp-1}\right|,
\end{equation}
and the squared amplitude equals 
\begin{equation}
A^2=\frac{\Gamma^2(\nu)}{4\pi^3}\left(\frac{1}{2}\frac{p}{mp-1}\right)^{1-2\nu}H_T^2\cdot
	\left(\frac{H_T}{M}\right)^{-2(n-1)\nu}.
\end{equation}
Note that, in the last equations, $k_T$ is the mode that crosses the Hubble radius at time $\eta_T$, $k_T=H_T$.  Since, by construction, the equation of motion of the field  $\varphi$ is linear, and because its modes are in a vacuum state, the fluctuations in $\varphi$ are Gaussian.

The primordial spectrum seeded during a stage of de Sitter inflation ($p=-1$) happens to be scale invariant for any value of $n$ if $m=n-1$. This includes the conventional dispersion relation, $n=1$ and $m=0$, and the dispersion relation of the ghost condensate \cite{Arkani-Hamed:2003uz}, $n=2$ and $m=1$. But as we see, it is also  possible to seed an exactly or nearly scale  invariant spectrum of primordial perturbations during a non-inflationary stage of expansion. In particular, many combinations of parameters lead to $n_s\approx 1$ in a non-accelerating universe ($p>1/2$), given that the only restriction on the model parameters so far is  $mp>1$, from equation (\ref{eq:condition}).  And even for $n=1$ the spectrum seeded during a non-inflationary phase can be scale-invariant if the effective speed of sound is time-dependent and decreasing ($m>0$), although this realization  requires a superluminal \emph{effective} speed of sound \cite{Armendariz-Picon:2003ht}.

To conclude this section, let us note that current observations actually disfavor an exact scale-invariant spectrum \cite{Spergel:2006hy} (see however \cite{Magueijo:2006we}). This does not pose any particular problem in our approach. Although in the realizations we discuss below $n$ has to be an integer, $p$ and $m$ can be arbitrary real numbers. In this case, just as for inflation, departures from scale invariance arise from, e.g., small departures of $p$ from integer values. 

\subsection{The scale M}
Within a  classical description of the universe,  nothing prevents us from taking the early-time limit $\eta\to 0$. However,  we expect quantum effects at Planckian energy densities to invalidate our semi-classical description. Hence, rather than requiring that modes be  adiabatic as $\eta\to 0$, we should require that cosmologically relevant modes be in the adiabatic vacuum at the Planckian time at the earliest.

Let us assume that the seeding stage (\ref{eq:a}) is followed by the conventional period of radiation domination, and let us assume that the transition occurs at time $\eta_T$. Requiring the adiabatic condition $\omega \, \eta\gg1 $ to hold at the Planck time for a mode of the size of our present horizon, we arrive at  the upper bound
\begin{equation}\label{eq:long}
	\left(\frac{M}{M_P}\right)^{n-1}\ll 10^{-29 n}
	\left(\frac{H_T}{M_P}\right)^{n/2-p(m+1)/(p+1)},
\end{equation} 
where $M_P=G^{-1/2}=10^{19}$~GeV is the Planck mass.  We shall later argue that the perturbations in the field $\varphi$ are converted into adiabatic perturbations at the time of the transition between the seeding stage and radiation domination. Because in our calculation of the spectrum we assumed that the relevant modes are long-wavelength, we have to make sure that this condition is satisfied at the time of the transition. Conservatively requiring than five decades in $k$ above our present horizon be long-wavelength at that time we arrive then at the lower bound 
\begin{equation}\label{eq:short} 
	\left(\frac{H_T}{M_P}\right)^{n/2-1}\ll \left(\frac{M}{M_P}\right)^{n-1} 10^{24n}. 
\end{equation}
Note that the universe has to be radiation dominated by the time of nucleosynthesis at the latest, which implies the additional constraint $H_T >10^{-44} M_P$.
Conditions (\ref{eq:long}) and (\ref{eq:short}) can be simultaneously satisfied with parameters that lead to a nearly scale invariant spectrum. Set for instance $p\approx 2$, $n=1$ and  $m=1$.  In this case, the mass scale $M$ does not enter the equations. Choose in addition $H_T=10^{-40} M_P$, which  amounts to a temperature at the transition $T_T\sim100$ MeV. All the required conditions are then met. If the parameters are  instead, say, $p\approx 1$, $n=3$ and $m=2$, the spectrum is again nearly scale invariant, and equations (\ref{eq:long}) and (\ref{eq:short}) are satisfied  with $M=10^{-17}$ eV and $H_T=10^{-40} M_P$. Note the  tiny value of the scale $M$, which is a reflection  of the horizon problem; in a non-accelerating universe, our current horizon has a size much larger than the Hubble radius at early times. However, the value of this parameter should be interpreted with care, as the scale $M$ is only defined modulo an appropriately normalized time-dependent function (see below). 

\section{Realizations}\label{sec:realizations}

We still have to answer whether the dispersion relation (\ref{eq:dispersion}) follows from any generally covariant scalar field theory.  What we have to explain is $i)$ the origin of the higher powers in spatial momentum $k$, and $ii)$ why these powers are not accompanied by corresponding powers of the scale factor $a$.  As we shall see,  ingredient $i)$ arises in any low-energy effective theory, while ingredient $ii)$ follows from couplings between $\varphi$ and  background scalars and tensors. 

\subsection{Effective Field Theories}
Let us begin by addressing the origin of the higher powers of spatial momenta in equation  (\ref{eq:dispersion}). Suppose we deal with  a low energy effective action with cut-off scale $M$. The action is expected to contain all possible terms compatible with the symmetries of the theory. Generically no symmetry forbids a term
\begin{equation}\label{eq:higher-order}
	L_\mathrm{corr}=\frac{1}{2 M^{2n-2}}\varphi \, \Box^n \varphi,
\end{equation}  
where $\Box=\nabla^\mu \nabla_\mu$, so we expect the action to contain this operator, in addition to many other similar ones. Upon variation of the action, this term leads to corrections to the equation of motion, $\Box\, \varphi+M^{2-2n}\Box^n \varphi=0$. In a spatially flat Friedman-Robertson-Walker universe with metric (\ref{eq:a}) these corrections have the structure
\begin{equation}
	\Box^n \varphi=\frac{1}{a^{2n+1}}\sum_{p+2q+s=2n} c^n{}_{pq} \cdot
	\eta^{-p} \, k^{2q} \,  v_k^{(s)},
\end{equation}
where $c^n{}_{pq}$ are dimensionless coefficients, a superscript $(s)$ denotes $s$ derivatives with respect to conformal time and we have reintroduced $v_k=a\,\varphi_k$. 

Although the equation of motion then contains the desired term, proportional to $k^{2n}$, it includes in addition many others that actually dominate over the gradients. Compare for instance $k^{2n}\,v_k$ with $(k^{2n-2}/\eta^2) v_k$. The first one dominates when $k^2>\eta^{-2},$ that is, for sub-horizon modes. But without inflation, cosmologically relevant scales are well outside the horizon in the past! Hence, higher-dimensional operators like the one in equation (\ref{eq:higher-order}) do not lead to the dispersion relation (\ref{eq:dispersion}).  

\subsection{Spatial Projectors}
Upon closer inspection of the structure of the dispersion relation (\ref{eq:dispersion}) one realizes that the key point is the presence of higher-order spatial derivatives in the equation of motion, but the absence at the same time of higher-order time derivatives. This structure naturally appears in the presence of appropriate second rank background tensors \cite{Jacobson:2000gw}. 

Imagine a spacetime endowed with a symmetric tensor $h_{\mu\nu}$ that projects onto the space orthogonal to a timelike vector field $u^\mu$.  By definition, the projector $h_{\mu\nu}$ satisfies $h_{\mu\nu}h^\nu{}_\rho=h_{\mu\rho}$ and $h_{\mu\nu}u^\nu=0$. In our context, the reader can think of the vector field $u^\mu$ as  the four velocity of observers for which the universe appears to be isotropic. In conformal time coordinates, this vector has components $u^\mu=(a^{-1},0,0,0)$, so the projection tensor is
$h_{\mu\nu}=a^2\cdot \mathrm{diag}(0,1,1,1)$. If this is the case, it then follows that, acting on a scalar,
\begin{equation}
	D_\mu = h_\mu{}^\nu \nabla_\nu
\end{equation}
is orthogonal to $u^\mu$ and has only spatial components, i.~e., is a spatial derivative. On the other hand
\begin{equation}
	d_\mu = (\delta_\mu{}^\nu-h_\mu{}^\nu) \nabla_\nu
\end{equation}
has vanishing spatial components and hence represents a time derivative. Therefore, by combining these two generally  covariant derivatives, we can obtain any combination of time and spatial derivatives. In particular, 
\begin{equation}\label{eq:spatial}
	\varphi \cdot (D_\mu D^\mu)^n  \varphi=(-1)^n  
	\frac{k^{2n}}{a^{2n}}\varphi_k^2.
\end{equation}

\subsection{Time functions}
Note that in equation (\ref{eq:spatial})  every factor of $k$ is accompanied by a corresponding factor of the scale factor $a$. In order to justify the presence of different powers of $a$ in the dispersion relation (\ref{eq:dispersion}),  a non-vanishing background scalar $r$ is sufficient. Because of the symmetry of the spacetime, this scalar can depend only on time, $r=r(\eta)$,  so this is why we shall call $r$ a time function. At this point let us simply  assume  that $r=a^{2(n-m-1)}$. We shall justify this functional dependence below; what matters here is that at time $\eta_T$, $r=1$.  Given equation (\ref{eq:a}), it follows that the Lagrangian
\begin{equation}\label{eq:L}
	L= -\frac{1}{2}d_\mu \varphi\,  d^\mu\varphi -\frac{(-1)^n}{2 M^{2n-2}} \cdot r \cdot 
	\varphi  \cdot (D_{\mu}D^{\mu})^n \varphi
\end{equation}
yields the equation of motion (\ref{eq:motion}) with dispersion relation given by (\ref{eq:dispersion}), provided  we assume that the fluctuations in $r$,  $g_{\mu\nu}$ and $h_{\mu\nu}$ are negligible; in other words, provided  that the universe is highly homogeneous and isotropic. Notice that $n=m+1$ leads to a simpler model, since in that case $r$ is a constant.    In the following, we present possible different origins of the spatial tensor $h_{\mu\nu}$ and the time function $r$. Similar constructions of the projection tensor  involving vector fields \cite{Jacobson:2000gw} are also possible.

\paragraph{Non-minimal gravitational couplings}
The existence of an appropriate tensor $h_{\mu\nu}$ is quite natural in a Friedman-Robertson-Walker universe, and does not imply the breaking of any spacetime symmetry. Consider for instance
\begin{equation}
	h_{\mu\nu}=\frac{3(p-1)}{p+1} \frac{R_{\mu\nu}}{R}
	+\frac{3g_{\mu\nu}}{2(p+1)} \;\text{and} \;
	r= \left(\frac{R}{R_T}\right)^{-\frac{p(n-m-1)}{p+1}}
\end{equation}
when $p\neq 1$.  The reader can readily verify that the tensor $h_{\mu\nu}$ and the scalar $r$ have the required properties in a universe whose scale factor evolves according to equation (\ref{eq:a}). This choice of $r$ defines $\eta_T$ to be the time at which the the Ricci scalar equals the free parameter $R_T$. The presence of the additional curvature terms however leads to  modifications in the  gravitational equations. Inverse powers of the Ricci scalar have been also postulated as explanations for cosmic acceleration in \cite{Carroll:2003wy}. Note that the scalar curvature does not necessarily vanish during radiation-domination, say, due to the presence of dark matter particles.

\paragraph{Couplings to other fields}
Even though in our analysis all we had to assume is the appropriate form of the scalar field equation of motion, regardless of the dynamics of gravity,  it might be phenomenologically desirable to restrict ourselves to actions where gravity is described by general relativity. If $\chi$ is a homogeneous time-evolving scalar field in a FRW universe, then, the
projector
\begin{equation}
	h_{\mu\nu}=g_{\mu\nu}-\frac{\partial_\mu\chi \,
	\partial_\nu \chi}{\partial_\alpha \chi \partial^\alpha \chi}
\end{equation}
also has the required properties. In addition, if the scalar field $\chi$ is rolling down an exponential potential
\begin{equation}\label{eq:chi}
	L_\chi=-\frac{1}{2}\partial_\mu \chi \partial^\mu \chi - V_0 \exp\left(-\sqrt{\frac{16\pi (1+p)}{p}}\frac{\chi}{M_P}\right),
\end{equation}
the time-keeping function
\begin{equation}
	r=\exp\left[(n-m-1)\sqrt{\frac{16\pi p}{1+p}}\frac{\chi-\chi_T}{M_P}\right]
\end{equation}
leads to  the desired power of the scale factor in the dispersion relation.\footnote{Since $\varphi$ vanishes classically, $\varphi_\mathrm{cl}\equiv \langle\varphi\rangle=0$, the terms in the Lagrangian (\ref{eq:L}) do not alter the $\chi$ equation of motion.} In this case, $\eta_T$ is the time at which the scalar field $\chi$ equals $\chi_T$. 

\section{Metric Perturbations}
The mechanism we have presented generates a nearly scale invariant spectrum of perturbations in a subdominant scalar field $\varphi$, that is, it leads to  \emph{isocurvature} perturbations. Observations however suggest that primordial perturbations are not only nearly scale invariant, but also \emph{adiabatic} and Gaussian. Hence, we still have to explain how these  isocurvature perturbations are converted into adiabatic  ones. We shall discuss this conversion in the context of the realization of our model where the spatial projectors arises from an evolving and homogeneous scalar field $\chi$. 

Imagine that the early universe is dominated by a scalar field  $\chi$ with an exponential potential, as in equation (\ref{eq:chi}). Assume that this potential has a minimum at $\chi_\mathrm{min}$, that is, the potential deviates from an exponential around $\chi\sim \chi_\mathrm{min}.$ Once the field reaches this minimum, it will start to oscillate and decay into the fields $\psi$ it couples to, thus reheating the universe.  Now, assume that the couplings of $\chi$ to $\psi$ are not constants, but depend on the fluctuating field $\varphi$ \cite{Dvali:2003em},
\begin{equation}
	L_\mathrm{int}=-\lambda_0 \cdot\left(1+f \frac{\varphi}{M}\right)\chi \bar{\psi}{\psi},
\end{equation}
where $\lambda_0$ and $f$ are two dimensionless constants. Because the reheating temperature depends on the coupling of $\chi$ to matter,  the fluctuations in $\varphi$  lead then to fluctuations in the reheating temperature and the Newtonian potential $\Phi$ \cite{Dvali:2003em, Armendariz-Picon:2003ku},
\begin{equation}
	\Phi\sim \frac{\delta T}{T}\sim f \frac{\varphi}{M}.
\end{equation}
Combining this equation with our results in equation (\ref{eq:power}) we find that in cases where the spectrum of $\varphi$ is nearly scale invariant,  the decay of $\chi$ results again in a nearly scale independent spectrum of adiabatic perturbations with squared  amplitude
\begin{equation}
	\mathcal{P}_\Phi\sim f^2 \cdot \left(\frac{H_T}{M}\right)^{\frac{3-n}{n}}.
\end{equation}
These perturbations are nearly Gaussian ($f_{NL}$ equals a few), provided that the decay rate of $\chi$ at the end of the seeding stage is not too high \cite{Zaldarriaga:2003my}.

Cosmic microwave background measurements \cite{Spergel:2006hy} require $\mathcal{P}_\Phi\sim 10^{-10}$, which fixes the parameter $f$ for given $H_T$ and $M$.  Hence, the amplitude of metric perturbations is not uniquely determined by the amplitude of the  perturbations in $\varphi$. For instance, for $n=3$ the amplitude of primordial perturbations is just fixed by $f$, and observations hence require $f\sim 10^{-5}$.

\section{Homogeneity}
Some inflationary models not only explain the origin of a nearly scale invariant spectrum of primordial perturbations, but also why the universe is homogeneous, isotropic and spatially flat.  
Whereas our scenario has nothing to say about the spatial geometry of the universe, it can indeed partially address the homogeneity of space.

To that purpose, consider the energy density of an inhomogeneous  field $\varphi$ in an expanding universe. The energy momentum tensor of the scalar field has been calculated in \cite{Lemoine:2001ar}. In the case of a scalar field with action (\ref{eq:L}) the energy density  due to the inhomogeneities in $\varphi$ is
 \begin{equation}
 	\rho_\mathrm{inh}\approx \frac{1}{2}\frac{\varphi_k'^2}{a^2}+\frac{1}{2}\frac{\omega^2 }{a^2} \varphi_k^2.
 \end{equation}
We would like to compare the energy density of these inhomogeneities with the total energy density of the universe, $\rho_\mathrm{tot} \sim M_P^2/ (a^2 \eta^2)$. We have seen that as the universe expands, modes leave the short-wavelength regime and become long-wavelength, that is $\omega \, \eta$ decreases. In the limit $\omega \, \eta\ll 1$, the solution of the equation of motion is $\varphi_k =const$. Therefore, the ratio of energy densities approaches
\begin{equation}
	\frac{\rho_\mathrm{inh}}{\rho_\mathrm{tot}}\sim \omega^2\, \eta^2 \, \frac{\varphi_k^2}{M_P^2}.
\end{equation}
and, hence, just as during inflation, decreases. Clearly, this argument is somewhat heuristic, but it nevertheless suggests that inhomogeneities are  ``washed-out" by the expansion even if the universe does not inflate. 

\section{Conclusions}
In this article we have described a way to seed a nearly scale invariant spectrum of adiabatic primordial perturbations in a non-inflating universe. The key element is the modified dispersion relation of a matter component in the universe.  No epoch of contraction is required, and the Hubble parameter does not have to increase at any time during the past history of the universe.  For concreteness, we have illustrated this idea with a scalar field, but other realizations  of this mechanism might be also possible. In this scalar field context, the modified dispersion relations arises from rather non-generic couplings of the scalar field to background scalars and tensors. Typically, these couplings have to be suppressed by a tiny scale $M$, a reflection of the horizon problem. Because the spectrum of primordial perturbations is seeded during a non-inflationary stage in an otherwise perfectly homogeneous universe, no gravitational waves are produced.  Perturbations are also expected to be non-Gaussian, with an $f_{NL}$ of a few, and thus potentially detectable in the near future. The mechanism does not address the spatial flatness of the universe, but might have the potential to tackle the homogeneity problem.

It is natural to compare our scenario with the  inflationary benchmark. Certainly, \emph{some} inflationary models still provide the simplest explanation for many of the features of our universe. But this work should be interpreted from a different perspective anyway. At this stage, it merely shows that the causal generation of  the primordial perturbations we observe  in our universe does not necessarily require a stage of inflation or an unconventional expansion history.

\begin{acknowledgments}
The author thanks the Yukawa Institute for Theoretical Physics and the Max-Planck-Institut f\"ur Gravitationsphysik for their kind hospitality. I am also indebted to Shinji Mukohyama for useful conversations and remarks. This work was supported in part by the NSF under grant PHY-0604760.
\end{acknowledgments}

\end{document}